\documentclass[conference]{IEEEtran}
\IEEEoverridecommandlockouts
\usepackage{textcomp}

\usepackage{cite}
\usepackage{amsmath,amssymb,amsfonts}
\usepackage{algorithmic}
\usepackage{graphicx}
\usepackage{optidef}
\usepackage{enumitem} % 用于自定义 enumerate 环境
\usepackage{float}
\usepackage{svg}  % For handling SVG files
\usepackage{float} % 在导言区添加这个包
\usepackage{placeins} % 用于插入 \FloatBarrier

\usepackage{textcomp}
\usepackage{xcolor}
\usepackage{subcaption}
\usepackage{bm}
\usepackage[
  letterpaper,
  left=0.75in,
  right=0.75in,
  top=0.75in,
  bottom=1.05in
]{geometry}

\ifCLASSINFOpdf
\else
\fi
\begin{document}
%
% paper title
% Titles are generally capitalized except for words such as a, an, and, as,
% at, but, by, for, in, nor, of, on, or, the, to and up, which are usually
% not capitalized unless they are the first or last word of the title.
% Linebreaks \\ can be used within to get better formatting as desired.
% Do not put math or special symbols in the title.
\title{Agentic Link Construction for Environment and Intent Aware 6G Communication}
%
%
% author names and IEEE memberships
% note positions of commas and nonbreaking spaces ( ~ ) LaTeX will not break
% a structure at a ~ so this keeps an author's name from being broken across
% two lines.
% use \thanks{} to gain access to the first footnote area
% a separate \thanks must be used for each paragraph as LaTeX2e's \thanks
% was not built to handle multiple paragraphs
%

\author{
    \IEEEauthorblockN{Zhaoyang Li, Shangzhuo Xie, Qianqian Yang\textsuperscript{\textsection}}
    \IEEEauthorblockA{College of Information Science and Electronic Engineering, Zhejiang University, Hangzhou, China}
    
    \IEEEauthorblockA{\{zhaoyangli, sz\_xie, qianqianyang20\}@zju.edu.cn}
}

\maketitle

% As a general rule, do not put math, special symbols or citations
% in the abstract or keywords.
\begin{abstract}
The emergence of sixth-generation (6G) networks heralds an intelligent communication ecosystem driven by the rapid proliferation of intelligent services and increasingly complex communication scenarios. However, current physical-layer designs—typically following modular and isolated optimization paradigms—fail to achieve global end-to-end optimality due to neglected inter-module dependencies. Although large language models (LLMs) have recently been applied to communication tasks such as beam prediction and resource allocation, existing studies remain limited to single-task or single-modality scenarios and lack the ability to jointly reason over communication states and user intents for personalized strategy adaptation. To address these limitations, this paper proposes a novel multimodal communication decision-making model for link construction leveraging reinforcement learning on pretrained LLMs. The proposed model semantically aligns channel state information (CSI) and textual user instructions, enabling comprehensive understanding of both physical-layer conditions and communication intents. It then generates physically realizable, user-customized link construction that dynamically adapts to changing environments and preference tendencies. A two-stage reinforcement learning framework is employed: the first stage expands the experience pool via heuristic exploration and behavior cloning to obtain a near-optimal initialization, while the second stage fine-tunes the model through multi-objective reinforcement learning considering BER, throughput, and power consumption. Experimental results demonstrate that the proposed model significantly outperforms conventional planning-based algorithms under challenging channel conditions, achieving robust, efficient, and personalized end-to-end communication strategies. Our dataset and code are available at https://github.com/LZhaoyang/AGENS.
\end{abstract}

% Note that keywords are not normally used for peerreview papers.

% For peer review papers, you can put extra information on the cover
% page as needed:
% \ifCLASSOPTIONpeerreview
% \begin{center} \bfseries EDICS Category: 3-BBND \end{center}
% \fi
%
% For peerreview papers, this IEEEtran command inserts a page break and
% creates the second title. It will be ignored for other modes.
\IEEEpeerreviewmaketitle

\section{Introduction}

The forthcoming sixth-generation (6G) wireless networks herald a paradigm shift that transcends conventional connectivity. Central to this vision is the concept of AI-native air interfaces, which enable self-optimizing networks capable of understanding user intent and dynamically adapting to complex and extreme environments\cite{b1,b2,b3}. However, current mainstream physical-layer architectures still rely on modular, layer-wise optimization frameworks, making it challenging to achieve the desired level of intelligence and scene-adaptivity\cite{b4}. Conventional designs typically decompose the system into separate, scene-irrelevant functional modules optimized in isolation\cite{b4}: advanced channel coding schemes such as LDPC and Polar codes are tailored to specific channel models\cite{b5}; adaptive modulation and coding (AMC) algorithms adjust modulation orders based on channel state information (CSI)\cite{b6}; and MIMO precoding and beamforming techniques aim to maximize spectral efficiency\cite{b7}. Although these methods enhance individual module performance, their isolated optimization neglects the complex, nonlinear interdependencies among components, resulting in globally suboptimal performance and diminished end-to-end efficiency.

With their powerful contextual understanding, cross-modal semantic fusion, and global reasoning capabilities, large language models (LLMs) can simultaneously process heterogeneous information and generate globally consistent decisions, offering new possibilities for overcoming the fragmented optimization inherent in traditional communication system design\cite{b8}. Existing research applying LLMs to wireless communications can generally be categorized into two main directions.

The first line of research focuses on leveraging pretrained LLMs to enhance specific communication tasks. For example, \cite{b9} proposes an LLM-based downlink channel prediction method, designing task-specific embedding layers for the frequency and angular domains while keeping the LLM backbone frozen. In \cite{b10}, an LLM-based beam prediction approach is proposed that utilizes textual prompts and achieves superior robustness and generalization compared with conventional LSTM-based models. In \cite{b11}, LLMs are employed for resource allocation and, after fine-tuning on small-scale datasets, achieve performance comparable to advanced reinforcement learning algorithms. Our previous work \cite{b12} proposes a modality-aligned LLM for channel prediction, which effectively narrows the modality gap between CSI and linguistic knowledge to enhance prediction accuracy. 

The second research direction focuses on developing task-specific foundation models tailored for wireless communication. For example, \cite{b13} proposes a task-independent universal channel embedding base model for communication by predicting the content of masked channel patches during training. Similarly, \cite{b14} considers the three dimensions of time, space, and frequency, and obtains a multi-task base model covering communication perception through pre-training and multi-task adaptive fine-tuning. \cite{b15} proposes a unified self-supervised framework that combines contrastive learning and masked reconstruction, specifically designed for multi-task channel representation learning. Despite these advances, notable limitations remain. Most existing efforts still focus on single tasks or modalities, leaving the core strengths of foundation models—deep multimodal reasoning, cross-modal fusion, and unified sequential decision-making—underexplored. 
This isolated usage pattern prevents communication systems from achieving unified situational awareness and globally optimal end-to-end decisions.
% This isolated usage pattern prevents communication systems from achieving unified situational awareness and globally optimal end-to-end decisions, thereby constraining the full potential of LLMs in the development of intelligent 6G communications.

To address the aforementioned challenges, we propose a foundation model designed for interactive communication strategy customization with Proximal Policy Optimization (PPO) reinforcement learning framework named  AGENS, which enables the construction of end-to-end communication links that adapt to specific user preferences under varying channel conditions. The main contributions of this paper can be summarized as follows:
\begin{itemize}
    \item We propose a foundation model for scenario-adaptive communication strategy optimization, termed AGENS. The model possesses the capability to comprehensively perceive both CSI and user intents. By semantically integrating physical-layer conditions with natural language instructions, AGENS can generate customized and optimal link configuration strategies that align with the current channel dynamics and user preferences, thereby achieving personalized and adaptive optimization of communication systems.
    
    \item We propose a two-stage reinforcement learning framework. In the first stage, a heuristic exploration algorithm is employed to expand the reinforcement learning experience pool, followed by behavior cloning via Supervised Fine-Tuning (SFT) to obtain a near-optimal initialization. In the second stage, the model is fine-tuned through reinforcement learning to further improve decision quality.
     
    \item Experimental results demonstrate that the proposed AGENS model can adaptively construct robust and efficient link strategies tailored to diverse user requirements under various challenging channel conditions, significantly outperforming existing planning-based algorithms.

\end{itemize}

\section{System model}
\subsection{Full-link Communication Simulation System}
% \begin{figure}[htbp]
%     \centering
%     \includegraphics[width=0.49\textwidth]{communication system.png} 
%     \caption{Communication system.}
%     \label{fig1}
% \end{figure}
We consider a 3GPP-compliant physical-layer communication link\cite{b16}, as illustrated in Fig.~\ref{fig1}. The system consists of several functional modules, including  channel coding, coding rate selection, modulation scheme, precoding, power control, channel estimation, and equalization. Each module adopts a specific strategy denoted as $c_i$, and the combination of all selected strategies forms a complete physical-layer transmission chain $\bm{a} = \{a_1, a_2, \ldots, a_n\}$.

The end-to-end transmission process of the communication link can be abstractly modeled as:
\begin{equation}
    \hat{\bm{x}} = f(\bm{x}; \bm{a}, \bm{H}),
\end{equation}
where $\bm{x}$ denotes the transmitted data, $\hat{\bm{x}}$ represents the recovered data at the receiver, and $\bm{H}$ is the CSI.

Based on the relationship between transmitted and received data, the overall performance of the physical-layer system can be comprehensively evaluated in terms of BER, system rate, and power consumption.

\subsection{State Space Modeling}

The selection of strategy combinations across modules in the communication link can be modeled as a RL process. Unlike conventional multi-step RL problems, the considered scenario corresponds to a single-step decision process, where each transmission involves a one-shot optimization of all module strategies.

We define the decision process as $M = \{\bm{S}, \bm{A}, r\}$, where $\bm{S}$ denotes the state space, $\bm{A}$ denotes the action space, and $r: \bm{S} \times \bm{A} \rightarrow \mathbb{R}$ represents the reward function. 
The objective is therefore to learn an optimal policy $\pi: \bm{S} \rightarrow \bm{A}$ that maximizes the expected immediate reward:
\begin{equation}
    \max_{\pi \in \Pi} \; \mathbb{E}_{\bm{s} \sim S,\, \bm{a} \sim \pi(a|s)} \big[ r(\bm{s}, \bm{a}) \big].
\end{equation}
where $\bm{a} \in \bm{A}$ denotes the action and $\bm{s} \in \bm{S}$ denotes the state. For generality, $\pi(\bm{a}|\bm{s}) \in [0,1]$ denotes the probability that policy $\pi$ selects action $\bm{a}$ given state $\bm{s}$.

In our framework, the proposed  model receives both textual information and CSI as inputs, and outputs the optimal strategy for each module in the physical-layer communication chain.  Accordingly, the policy parameterized by $\theta$ can be expressed as:
\begin{equation}
    \pi_{\theta}: \mathcal{O}_{\text{text}} \times \mathcal{O}_{\text{CSI}} \rightarrow \bm{a},
\end{equation}
where $\mathcal{O}_{\text{text}}$ denotes the textual observation space, $\mathcal{O}_{\text{CSI}}$ represents the channel state information space, and $\bm{c}$ corresponds to the set of strategies for all physical-layer modules. 
% This formulation allows the proposed model to integrate semantic understanding from textual prompts and physical knowledge from CSI, enabling semantic-level reasoning and joint optimization of multi-module communication strategies.

\section{Proposed Method}
In this section, we introduce the  structure of our proposed AGENS and the training method.
\subsection{Multimodal Communication Decision Large Language Model}
\begin{figure*}[htbp]
    \centering
    \includegraphics[width=1\textwidth]{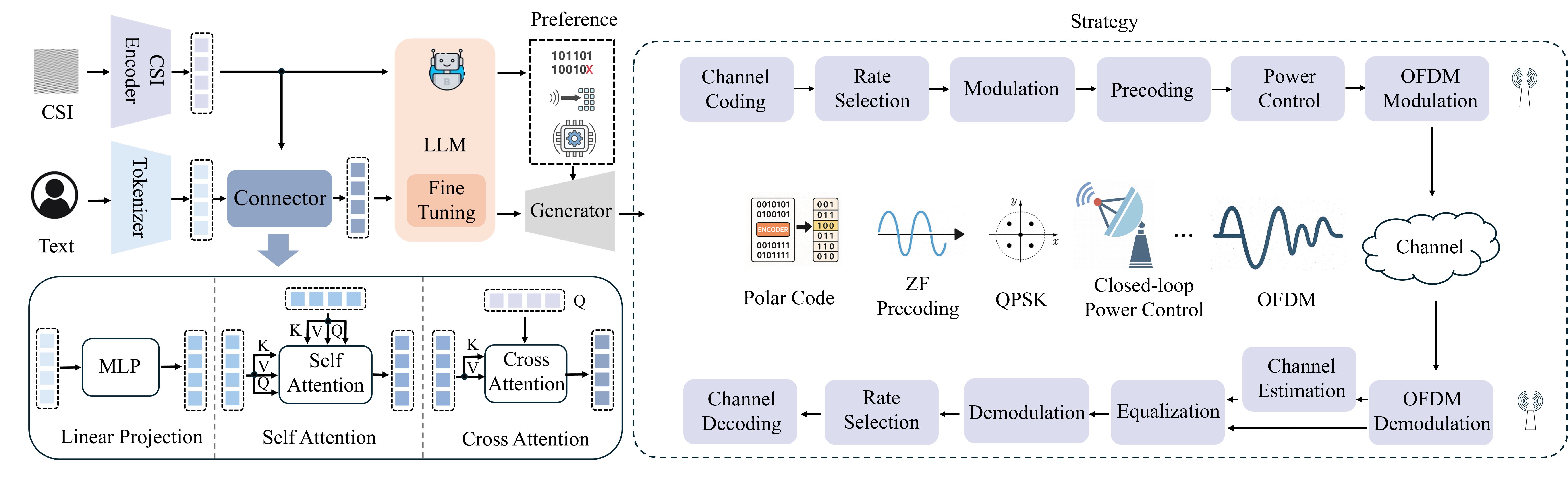} 
    \caption{Our proposed AGENS.}
    \label{fig1}
\end{figure*}

As illustrated in Fig.~\ref{fig1}, the AGENS consists of a text encoder, a CSI encoder, a cross-modal attention module for textual semantic compression, a pretrained LLM, and a generator responsible for final strategy generation.
The generator comprises $n$
 actor networks, each designed to produce the optimal configuration strategy for a specific physical-layer module.
% \begin{figure}[t]
%   \centering

%   \begin{subfigure}{0.49\textwidth}
%     \centering
%     \includegraphics[width=\linewidth]{llm.png}
%     \caption{Our proposed model.}
%   \end{subfigure}\vspace{0.8em}

%   \begin{subfigure}{0.49\textwidth}
%     \centering
%     \includegraphics[width=\linewidth]{system.png}
%     \caption{Communication system.}
%   \end{subfigure}

%   \caption{(a) Our proposed model, (b) Communication system.}
%   \label{fig1}
% \end{figure}
The input CSI sequence $\bm{H} \in\mathbb{R}^{L \times F}$ is processed by a CSI encoder to extract CSI features $\bm{{\mathit{Y}}}_{\text{csi}}$, where the CSI encoder adopts the same architecture as the preprocessing module in our previous work\cite{b11}, which can be expressed as:
\begin{equation}
        \bm{{\mathit{Y}}}_{\text{csi}} = \text{Encoder}_\text{csi} (\bm{H}) ,
\end{equation}where $\bm{{\mathit{Y}}}_{\text{csi}} \in\mathbb{R}^{L \times d}$.

Meanwhile, the user request $\bm{T}$ is converted into text tokens through a tokenizer and embedded into the same semantic space as the LLM. The embedding process can be formulated as:
\begin{equation}
    \bm{T}_{\text{text}} = \text{Embed}(\text{Tokenizer}(\bm{T})), 
\end{equation}where $\bm{T}_{\text{text}} \in\mathbb{R}^{L_1 \times d}$ represents the word embedding of the original text.

Directly feeding long text sequences into the LLM leads to excessive computational overhead. To address this issue, we introduce a connector module before the text tokens are input into the LLM, which performs semantic filtering on textual embeddings. As illustrated in Fig.~\ref{fig1}, the connector consists of three main stages: linear projection, self attention , and cross attention. First, a fully connected layer is employed to reduce the sequence length of text embeddings:
\begin{equation}
    \bm{Z}_{\text{text}} = \text{Linear}(\bm{T_{\text{text}}}) ,
\end{equation}
where $\bm{Z}_{\text{text}} \in \mathbb{R}^{L_2 \times d}$.

Next, a self-attention mechanism is applied to establish semantic dependencies and aggregate global textual information:
\begin{equation}
    \tilde{\bm{Z}}_{\text{text}} =
    \text{Self-Attention}
    ({\bm{Z}}_{\text{text}}),
\end{equation}

Finally, inspired by the Q-Former structure, a cross-modal attention mechanism is adopted, where the CSI features act as queries to filter the textual embeddings semantically. Let \(\bm{Y}_{\text{csi}} \in \mathbb{R}^{L \times d}\) denote the CSI representation aligned with the LLM’s semantic space. The cross-modal attention process can be formulated as:

\begin{equation}
\bm{{\mathit{Y}}}_{\text{text}}= \text{Softmax}\left(\frac{Q K^T}{\sqrt{c}}\right)V,    
\end{equation}where
\begin{equation}
Q = \bm{{\mathit{Y}}}_{\text{csi}}W_Q, \quad K = \tilde{\bm{Z}}_{\text{text}}W_K, \quad V = \tilde{\bm{Z}}_{\text{text}}W_V,    
\end{equation}
where $\mathbf{W}_Q$, $\mathbf{W}_K$, and $\mathbf{W}_V \in \mathbb{R}^{d \times d}$ are attention parameters. This operation allows CSI-driven queries to focus on task-relevant semantic tokens, effectively compressing long textual inputs into concise and context-aware embeddings.

After obtaining the CSI-aligned text features, we concatenate them with the projected CSI embeddings and feed them into the pretrained LLM backbone:
\begin{equation}
    \bm{{\mathit{Y}}}_{\text{LLM}} = \text{LLM}\left(\text{Concat}[\bm{{\mathit{Y}}}_{\text{text}}; \bm{{\mathit{Y}}}_{\text{csi}}]\right),
\end{equation}
where $\bm{{\mathit{Y}}}_{\text{LLM}} \in \mathbb{R}^{2L \times d}$ denotes the hidden representation from the final transformer layer. 

Finally, the multimodal feature $\bm{\mathit{Y}}_{\mathrm{LLM}}$ is fed into multiple executor networks to generate the final decision strategies for different physical-layer modules. The output of the $i$-th executor network can be expressed as:
\begin{equation}
    a_{i} = g_{\text{actor},i}\!\left(\bm{{\mathit{Y}}}_{\text{LLM}}\right), \quad i = 1, 2, \ldots, n,
\end{equation}
where $n$ denotes the number of decision modules such as channel coding, modulation, power control, and resource allocation. The overall physical-layer configuration is thus represented as:
\begin{equation}
\bm{a} = \{a_1, a_2, \ldots, a_n\},
\end{equation}
which forms the final multimodal communication strategy determined jointly by the semantic intent and the CSI conditions.

\subsection{Training Method}
\begin{figure}[htbp]
    \centering
    \includegraphics[width=0.49\textwidth]{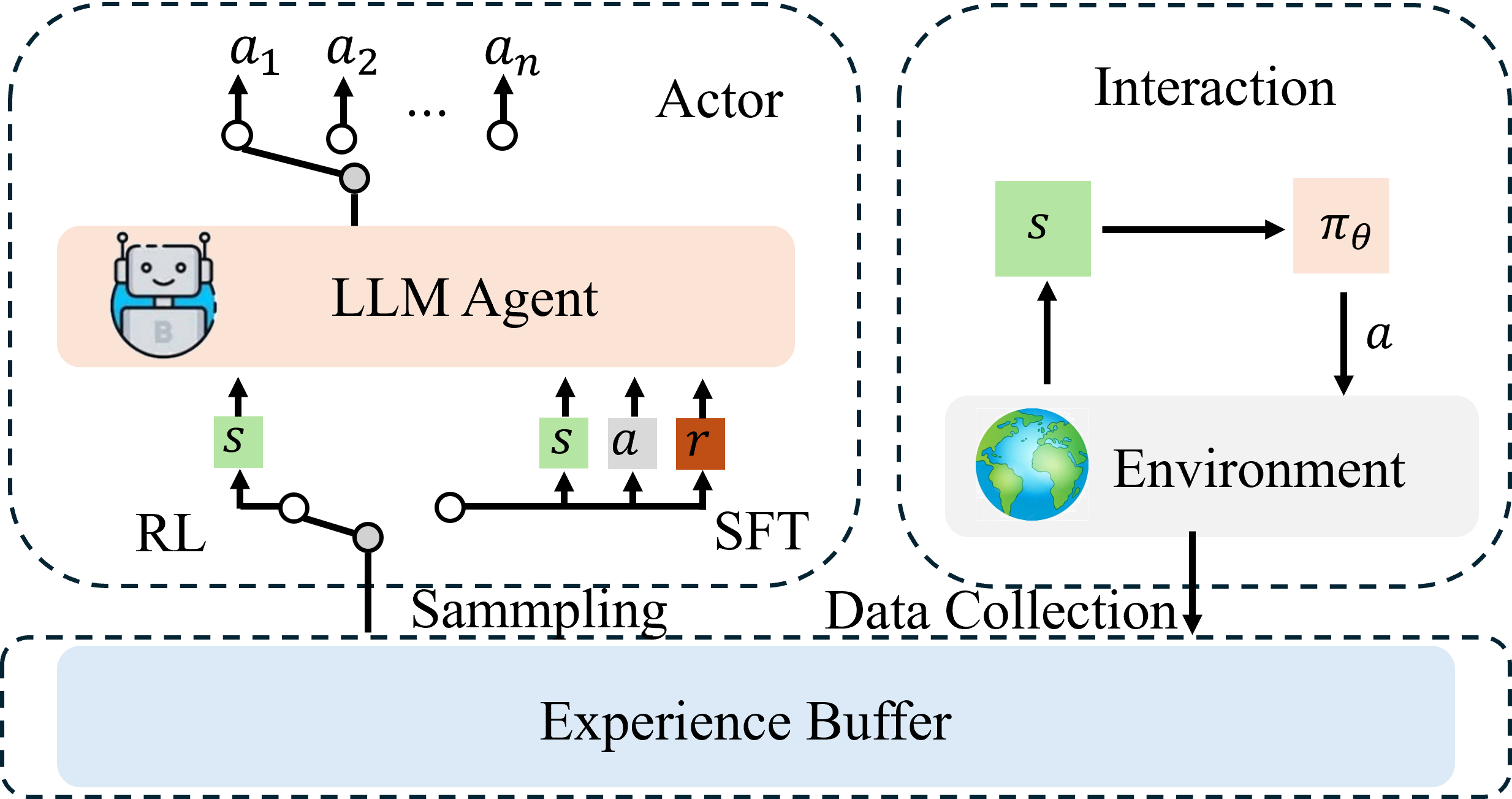} 
    \caption{Illustration of the proposed RL method.}
    \label{fig2}
\end{figure}
As illustrated in Fig.~\ref{fig2}, the proposed model is trained using a two-stage procedure that combines SFT and  reinforcement learning fine-tuning. The objective of this training strategy is to optimize the multimodal decision model so that it can effectively reason over both textual and CSI features to generate adaptive and high-performance communication strategies.

In the first stage, SFT is conducted using a set of high-quality decision samples generated by a greedy search algorithm. During early exploration, the multimodal model interacts with the environment to produce decision tuples $(\bm{\mathit{s}}, \bm{\mathit{a}}, r)$, where $\bm{\mathit{s}}$ denotes the state consisting of textual and CSI embeddings, $\bm{\mathit{a}}$ represents the selected communication strategy, and $r$ is the corresponding reward. The greedy search algorithm selects actions that maximize the immediate reward, and the resulting tuples are stored in a replay buffer $\mathcal{D}$. The model is then pretrained in a supervised manner to imitate these high-quality decisions by minimizing the discrepancy between the predicted actions and the stored actions in $\mathcal{D}$ :
\begin{equation}
    \mathcal{L}_{\text{SFT}} = - \mathbb{E}_{(\bm{{\mathit{s}}}, \bm{{\mathit{a}}}) \sim \mathcal{D}} \big[ \log \pi_{\theta}(\bm{{\mathit{a}}} | \bm{{\mathit{s}}}) \big],
\end{equation}
where $\pi_{\theta}$ denotes the policy parameterized by the LLM-based actor and $\theta$ represents the trainable parameters. This SFT stage enables the model to quickly learn a near-optimal policy by mimicking the expert trajectories in the buffer, effectively providing a warm start for subsequent reinforcement learning.

After SFT, the model undergoes reinforcement learning fine-tuning to further optimize decision performance under the true environment dynamics. Given the state $\bm{{\mathit{s}}}$, the model samples an action $\bm{{\mathit{a}}} \sim \pi_{\theta}(\bm{{\mathit{a}}}|\bm{{\mathit{s}}})$, receives a scalar reward $r = f_{\text{reward}}(\text{BER}, \text{Rate}, \text{Power})$, and updates its parameters to maximize the expected return:
\begin{equation}
    J(\theta) = \mathbb{E}_{\bm{{\mathit{s}}}, \bm{{\mathit{a}}} \sim \pi_{\theta}} [ r(\bm{{\mathit{s}}}, \bm{{\mathit{a}}}) ].
\end{equation}

The gradient of the objective can be estimated using the policy gradient method as
\begin{equation}
    \nabla_{\theta} J(\theta) = \mathbb{E} \big[ \nabla_{\theta} \log \pi_{\theta}(\bm{{\mathit{a}}}|\bm{{\mathit{s}}}) \cdot r(\bm{{\mathit{s}}}, \bm{{\mathit{a}}}) \big].
\end{equation}

% This formulation allows the model to retain the expert priors learned from the first stage while gradually improving through reinforcement signals.

The reward function is designed as a weighted linear combination of three key communication performance indicators: BER, Rate, and complexity. The overall reward is defined as
\begin{equation}
    r = w_{\text{ber}} \cdot R_{\text{ber}} + w_{\text{rate}} \cdot R_{\text{rate}} + w_{\text{power}} \cdot R_{\text{power}},
\end{equation}
where $w_{\text{ber}}, w_{\text{rate}}, w_{\text{power}} \in [0, 1]$ are weighting coefficients satisfying $w_{\text{ber}} + w_{\text{rate}} + w_{\text{power}} = 1$, and the normalized terms $R_{\text{ber}}$, $R_{\text{rate}}$, and $R_{\text{power}}$ represent the contributions of BER, rate, and power consumption, respectively.

\section{Numerical Results}
This section presents our simulation results. First, we describe the simulation setup in detail. Then, we compare AGENS  with several existing policy selection methods. In addition, we present a human-computer interactive question-answering example to demonstrate the intelligence and flexibility of AGENS.
\subsection{Experimental Setup}\label{AA}
We generate CSI datasets using Sionna-RT with diverse parameter configurations. For each CSI sample, a corresponding natural-language user-intent label is assigned to build diverse task scenarios. In the design of user intents, we consider three representative 6G service preferences: high throughput, high reliability, and energy-efficient operation. To evaluate the performance of the constructed communication link, we adopt BER, rate (defined as the ratio of the number of successfully transmitted bits to the number of original bits), and power compensation (i.e., the amount of transmit-side compensation power, measured in dB) as the evaluation metrics. For model implementation, GPT-2 Medium (355M parameters) is employed as the backbone architecture for building the AGENS. It should be noted that the pretrained LLM is mainly employed for its strong language understanding and user-intent inference capabilities. The proposed framework can be easily replaced with other pretrained language models.
\subsection{Model Decision Performance}\label{BB}
We first evaluate the capability of AGENS to generate personalized communication strategies under varying user preferences. For comparison, we consider four baseline methods: random strategy selection, greedy search, beam search (with a beam width of 2), and a Transformer-based variant of AGENS, where the LLM component is replaced by a standard Transformer. The Transformer-based baseline is introduced to highlight the effectiveness of incorporating LLMs within our framework. For the random strategy baseline, different random seeds are employed for different user preferences to ensure sufficient diversity in the generated strategies.
\begin{figure*}[t]
    \centering
    % 第一张图
    \begin{minipage}[t]{0.32\textwidth}
        \centering
        \includegraphics[width=\textwidth]{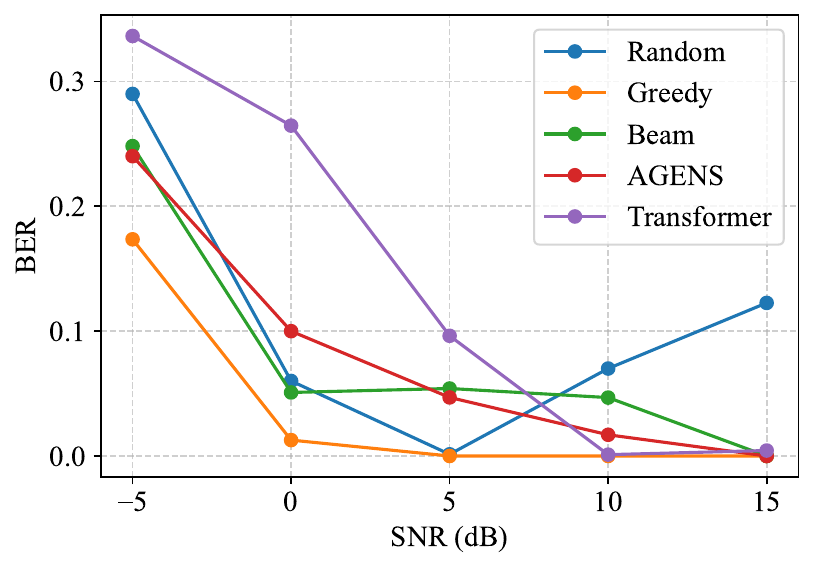}
        \caption*{(a)  BER }
    \end{minipage}
    \hfill
    % 第二张图
    \begin{minipage}[t]{0.32\textwidth}
        \centering
        \includegraphics[width=\textwidth]{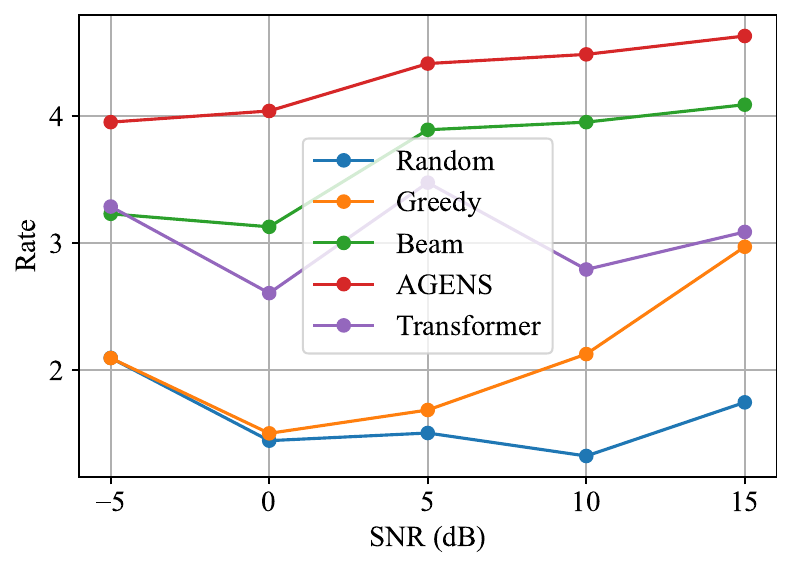}
        \caption*{(b)  Rate }
    \end{minipage}
    \hfill
    % 第三张图
    \begin{minipage}[t]{0.32\textwidth}
        \centering
        \includegraphics[width=\textwidth]{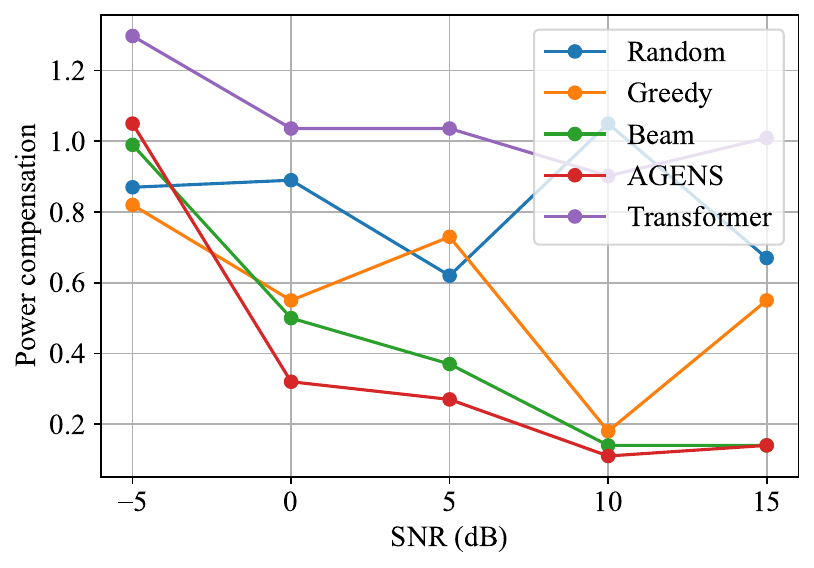}
        \caption*{(c)  Power Consumption }
    \end{minipage}
    \caption{Performance comparison of different strategy selection methods for high throughput strategy under various SNRs.}
    \label{fig3}
\end{figure*}

\begin{figure*}[t]
    \centering
    % 第一张图
    \begin{minipage}[t]{0.32\textwidth}
        \centering
        \includegraphics[width=\textwidth]{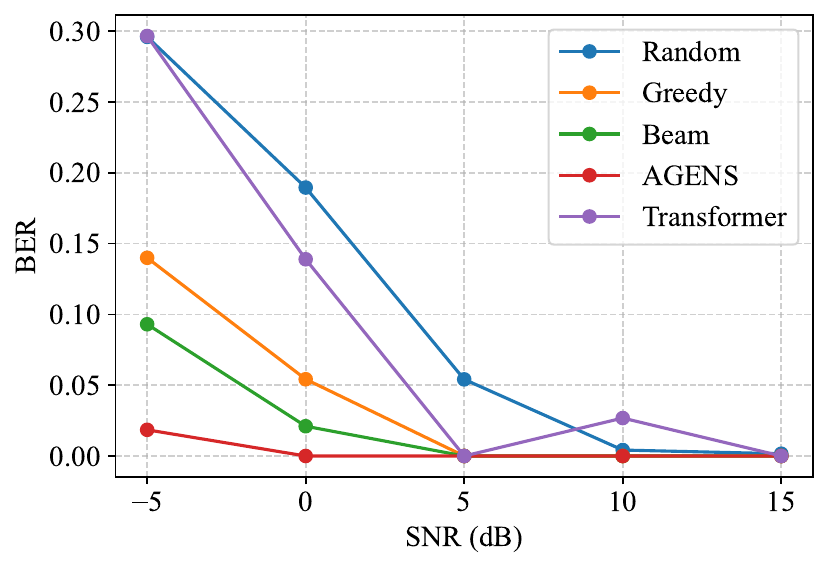}
        \caption*{(a)  BER }
    \end{minipage}
    \hfill
    % 第二张图
    \begin{minipage}[t]{0.32\textwidth}
        \centering
        \includegraphics[width=\textwidth]{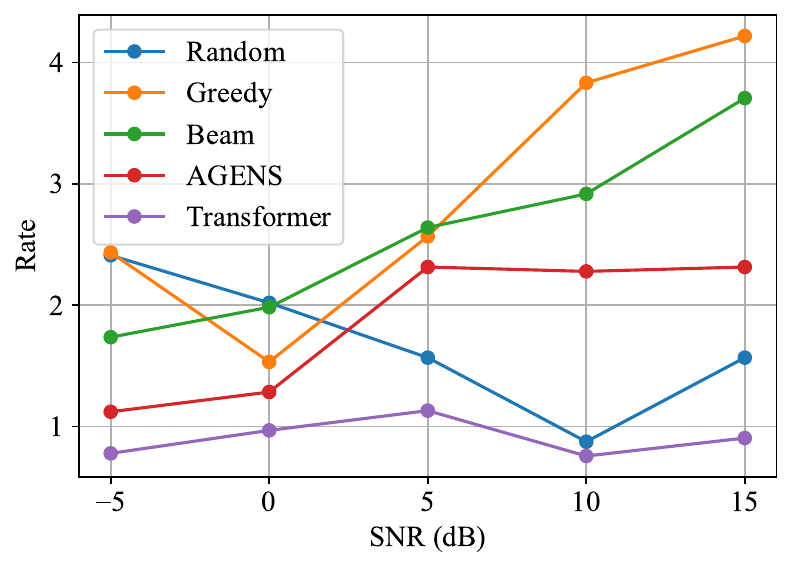}
        \caption*{(b)  Rate }
    \end{minipage}
    \hfill
    % 第三张图
    \begin{minipage}[t]{0.32\textwidth}
        \centering
        \includegraphics[width=\textwidth]{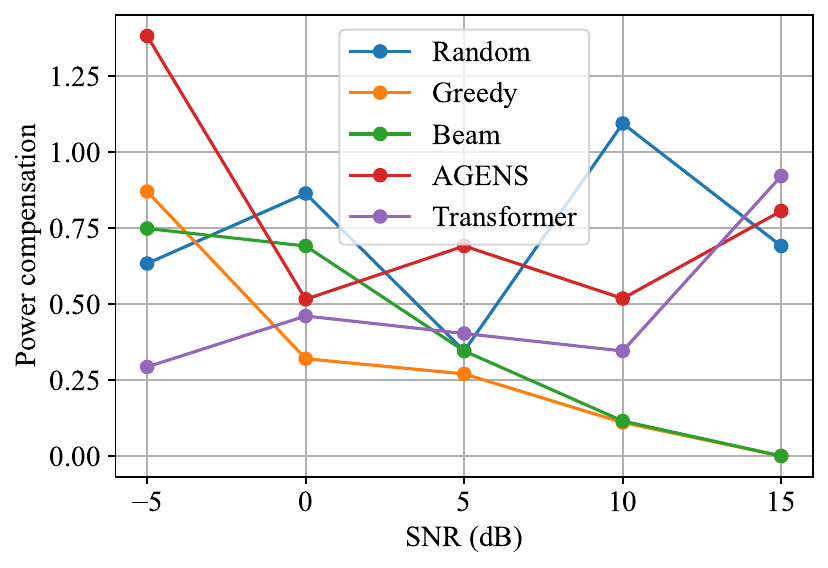}
        \caption*{(c)  Power Consumption }
    \end{minipage}
    \caption{Performance comparison of different strategy selection methods for high reliability strategy under various SNRs.}
    \label{fig4}
\end{figure*}

As shown in Fig.~\ref{fig3}, when high throughput is prioritized, all methods trade BER for higher transmission rates. Beam search achieves the best performance among heuristic baselines due to its larger search space, yet remains limited by local optimization. In contrast, AGENS attains higher rates while maintaining BER within an acceptable range, achieving a better balance between throughput and reliability. When reliability is emphasized, AGENS generates more robust strategies that better align with user intent. As illustrated in Fig.~\ref{fig4}, although both heuristic methods and AGENS produce low-BER strategies, AGENS shows clear advantages in low-SNR regimes. This is partly because it adopts lower transmission rates and higher transmit power, consistent with practical system design principles. Under the energy efficiency objective, beam search and AGENS achieve comparable performance. However, AGENS still attains slightly lower BER while maintaining low power consumption, indicating its superior global planning capability over heuristic methods.

\begin{figure*}[t]
    \centering
    % 第一张图
    \begin{minipage}[t]{0.32\textwidth}
        \centering
        \includegraphics[width=\textwidth]{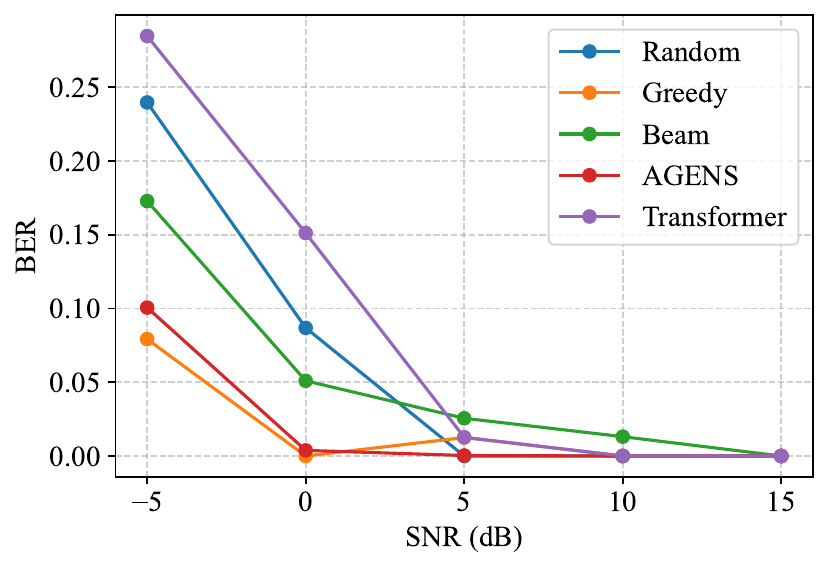}
        \caption*{(a) BER }
    \end{minipage}
    \hfill
    % 第二张图
    \begin{minipage}[t]{0.32\textwidth}
        \centering
        \includegraphics[width=\textwidth]{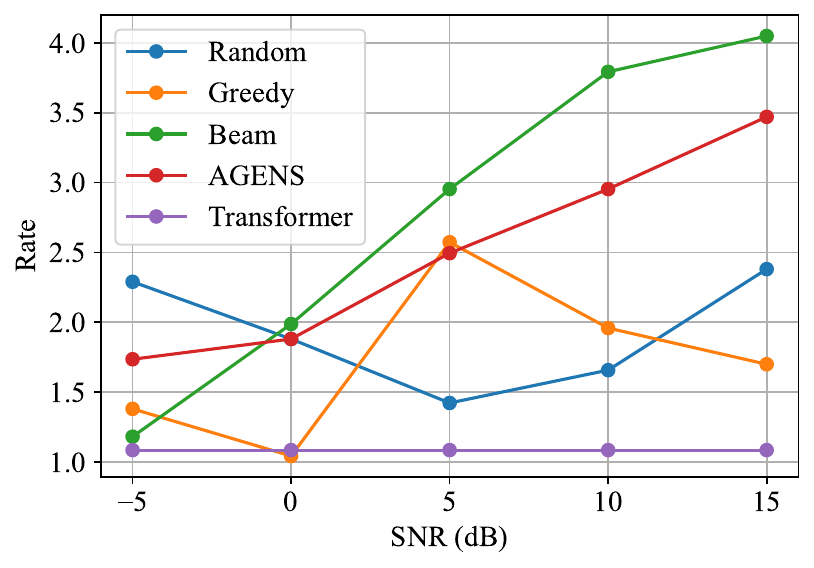}
        \caption*{(b)  Rate }
    \end{minipage}
    \hfill
    % 第三张图
    \begin{minipage}[t]{0.32\textwidth}
        \centering
        \includegraphics[width=\textwidth]{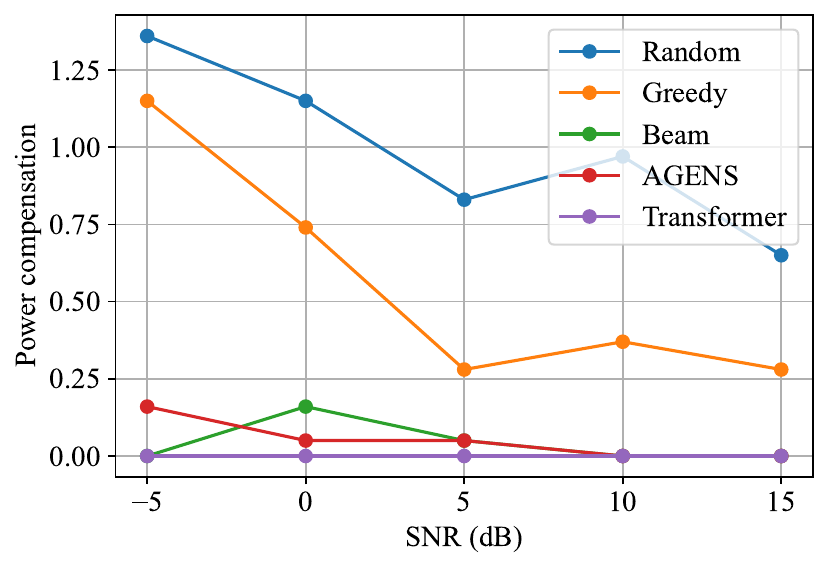}
        \caption*{(c) Power Consumption  }
    \end{minipage}
    \caption{Performance comparison of different strategy selection methods for energy-efficient strategy under various SNRs.}
    \label{fig5}
\end{figure*}

It can also be observed that the Transformer-based method performs worse than traditional heuristic approaches when either high throughput or high reliability is prioritized. Although it achieves relatively low power consumption under the energy efficiency objective, it often produces overly aggressive strategies. As shown in Fig.~\ref{fig5}, the Transformer-based method tends to generate rigid decisions, selecting nearly identical transmission rates across all scenarios. Such strategies result in excessively high BER, which may render the communication system unreliable in practice. These observations highlight the necessity of incorporating LLMs. With stronger reasoning and contextual understanding capabilities, LLMs enable more effective and adaptive communication strategy planning.

Finally, some performance curves exhibit noticeable fluctuations. This is partly due to the inherent randomness of the random strategy, and partly because heuristic methods rely on local optimization of individual modules, which may lead to unstable overall performance in multi-module communication systems. In contrast, AGENS produces smoother curves across different metrics, indicating better global coordination and higher stability.
\subsection{Model Complexity}\label{CC}
\begin{table}[ht]
\centering
\caption{Inference time.}
\label{TableI}
\begin{tabular}{|c|c|c|c|}
\hline

 \textbf{method} & \textbf{inference time} \\
\hline
  Random  & 0.005s \\
\hline
   AGENS & 0.167 s \\
\hline
  Beam  & 446.472s \\
\hline
  Greedy  & 307.987s \\
\hline

\end{tabular}
\end{table}
To evaluate the practical feasibility of deploying the proposed framework in real-world communication systems, we analyze the algorithmic complexity of different strategy planning methods for a single planning task. As shown in Table $\ref{TableI}$, heuristic-based methods such as greedy search and beam search exhibit significantly higher inference latency than AGENS due to their sequential search and evaluation processes, where candidate actions must be explicitly explored or optimized. In contrast, AGENS leverages end-to-end learning to capture the dependencies among modules and generate globally optimized strategies through a single forward pass. This greatly reduces inference time and demonstrates the scalability and practicality of the proposed model.

% In real-time physical layer link construction—where channel conditions and user requirements may change rapidly—such low-latency decision-making is critical. 
% In the ablation study, we compare the complete CoT-RL training framework with a variant that excludes the behavior cloning stage and relies solely on reinforcement learning. As illustrated in Fig.$\ref{fig7}$, when behavior cloning supervision is removed, the model struggles to maintain a proper balance between BER and throughput under the high-throughput preference setting. Consequently, the resulting strategies become overly conservative, leading to noticeably lower system throughput. These observations indicate that behavior cloning provides a crucial initialization that enables the model to learn effective preference-aware strategies. In contrast, pure reinforcement learning tends to converge to conservative policies that generalize more easily but fail to achieve optimal performance.
% \begin{figure*}[t]
%     \centering
%     % 第一张图
%     \begin{minipage}[t]{0.48\textwidth}
%         \centering
%         \includegraphics[width=\textwidth]{BER_ab.pdf}
%         \caption*{(a) BER }
%     \end{minipage}
%     \hfill
%     % 第二张图
%     \begin{minipage}[t]{0.48\textwidth}
%         \centering
%         \includegraphics[width=\textwidth]{Rate_ab.pdf}
%         \caption*{(b)  Rate }
%     \end{minipage}
%     \hfill
%     \caption{Ablation experiments.}
%     \label{fig7}
% \end{figure*}
\subsection{Model Interaction}\label{DD}
\begin{figure}[htbp]
    \centering
    \includegraphics[width=0.49\textwidth]{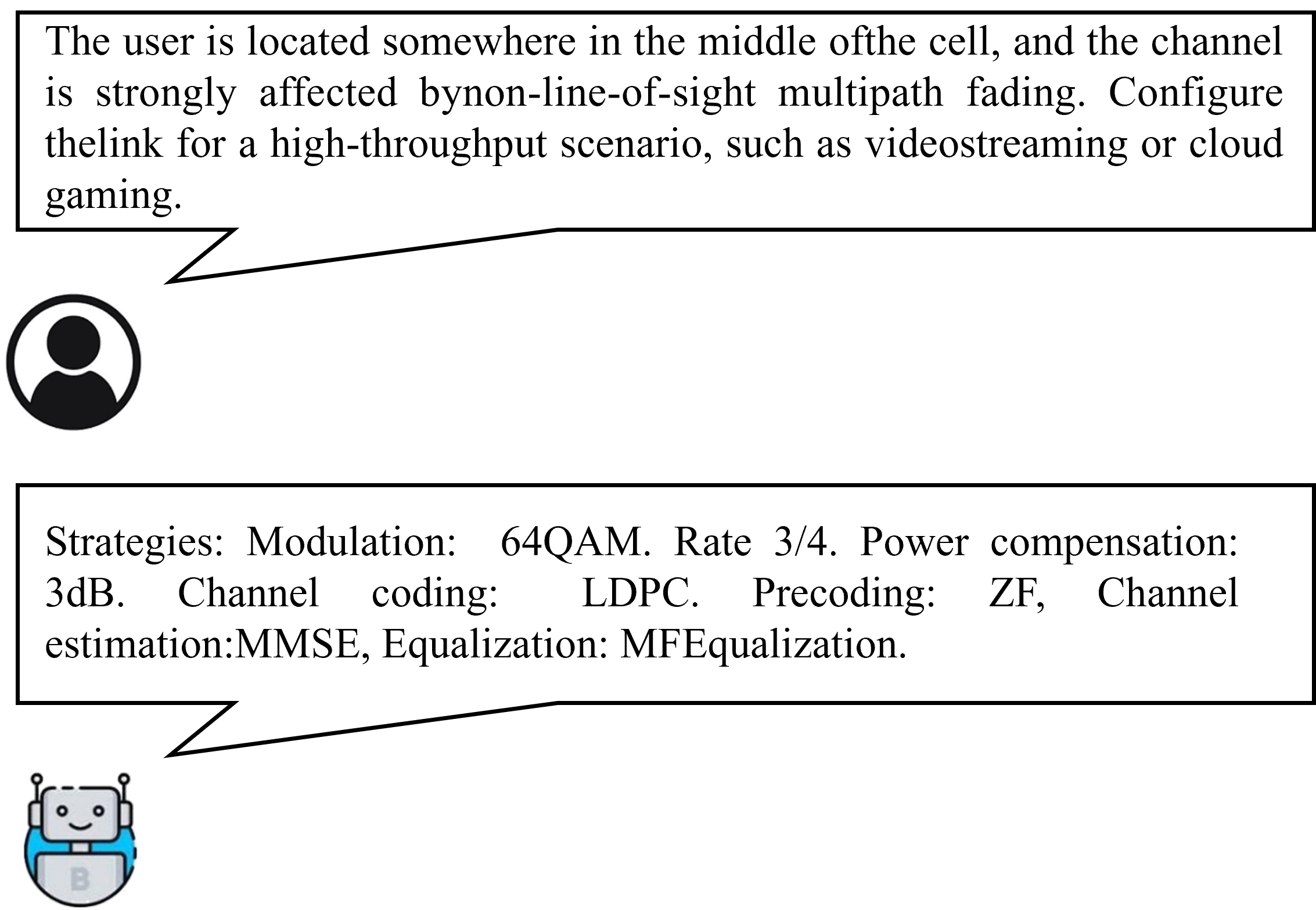} 
    \caption{An input-output example of AGENS.}
    \label{fig6}
\end{figure}
In this section, we present a human--machine interaction example to illustrate the capabilities of the proposed AGENS. As shown in Fig.~\ref{fig6}, under a given communication condition, the user provides environmental information and transmission intent as multimodal inputs to the model. Based on these inputs, AGENS interprets the user intent, understands the communication context, and infers the user preference. For example, from the query in Fig.~\ref{fig6}, the model identifies that high throughput is prioritized in the communication system. Accordingly, it generates an optimized and interpretable communication strategy. 
\section{Conclusion}
In this paper, we proposed AGENS, a foundation model for interactive communication strategy customization. AGENS has been designed to comprehensively perceive both CSI and user intent. By integrating physical-layer conditions with natural-language instructions, the model first generates an intermediate chain of thought to infer user-specific preferences, and subsequently produces a customized and optimal link configuration strategy that aligns with both the current channel state and the inferred user intent, thereby enabling personalized and adaptive optimization of communication systems. During training, heuristic algorithms have been employed to collect high-quality experience samples for SFT, followed by reinforcement learning fine-tuning to obtain the optimal policy. Extensive experimental evaluations have demonstrated that AGENS significantly outperforms traditional planning-based algorithms under various challenging channel conditions, while flexibly generating optimal physical-layer strategies tailored to different user preferences.

\end{document}